\documentclass[prl,aps,reprint,amsmath,amssymb,superscriptaddress,floatfix]{revtex4-1}

\usepackage[colorlinks,citecolor=blue,linkcolor=blue]{hyperref}
\usepackage{graphicx}% Include figure files
\usepackage{dcolumn}% Align table columns on decimal point
\usepackage{bm}% bold math

\begin{document}

\title{Spin wave power flow and caustics in ultrathin ferromagnets with the Dzyaloshinskii-Moriya interaction}

\author{Joo-Von Kim}
\email{joo-von.kim@u-psud.fr}
\affiliation{Centre for Nanoscience and Nanotechnology (C2N), CNRS, Univ. Paris-Sud, Universit{\'e} Paris-Saclay, 91405 Orsay, France}
\author{Robert L. Stamps}
\affiliation{SUPA School of Physics and Astronomy, University of Glasgow, Glasgow G12 8QQ, United Kingdom}
\author{Robert E. Camley}
\affiliation{Department of Physics and Energy Science, University of Colorado at Colorado Springs, 1420 Austin Bluffs Pkwy, Colorado Springs CO 80918, USA}

\date{\today}

\begin{abstract}
The Dzyaloshinskii-Moriya interaction in ultrathin ferromagnets can result in nonreciprocal propagation of spin waves. We examine theoretically how spin wave power flow is influenced by this interaction. We show that the combination of the dipole-dipole and Dzyaloshinskii-Moriya interactions can result in unidirectional caustic beams in the Damon-Eshbach geometry. Morever, self-generated interface patterns can also be induced from a point-source excitation.
\end{abstract}

\maketitle

%%
%	begin text
%%

The Dzyaloshinskii-Moriya interaction (DMI) is a short-range chiral spin-spin interaction in systems lacking inversion symmetry~\cite{Dzyaloshinsky:1958vq, Moriya:1960kc, Bogdanov:2001hr}. In ultrathin ferromagnetic metals, this interaction can be induced at an interface with a normal metal possessing a strong spin-orbit coupling~\cite{Fert:1980hr, Fert:1990}. The interfacial form has received significant attention in recent years, where among the highlights are the creation of skyrmions at room temperature~\cite{Jiang:2015cs, MoreauLuchaire:2016em, Boulle:2016jt, Woo:2016jw} and the fast current-driven motion of chiral domain walls~\cite{Ryu:2013dl, Emori:2013cl}. In terms of dynamic effects the DMI also introduces a nonreciprocity in spin wave propagation, where $\omega(k) \neq \omega(-k)$. This effect, first predicted and observed in epitaxial Fe/W layers~\cite{Udvardi:2009fm, Zakeri:2010ki}, has since been observed in other sputtered systems using Brillouin light scattering~\cite{Di:2015fe, Di:2015fs, Nembach:2015ep, Belmeguenai:2015hj, Cho:2015eq}.

However, one feature that has not been significantly investigated is the issue of power flow.  It is immediately clear that this is a requirement from the shifting of the spin wave dispersion curve introduced by DMI.  With DMI and for propagation perpendicular to the magnetization the dispersion curve is approximately a parabola but with the minimum shifted away from the origin along the wave vector axis.  Because of this $d\omega/dk$ is negative in some regions, and this indicates the group velocity is opposite to the phase velocity.  However, this simple analysis is not sufficient to capture all the important features of the anisotropic power flow created by the DMI. We note that the study of focusing patterns for bulk~\cite{Taylor:1969} and surface phonons~\cite{Camley:1983} in crystals is well known.  The corresponding investigations in thick film magnetic systems have begun only recently with both experimental~\cite{Demidov:2007cw, Demidov:2009gm, Schneider:2010ej, Sebastian:2013cu} and theoretical results~\cite{Veerakumar:2006gy}.  The focusing results have already shown remarkable behaviors, including focusing effects of energy well below the expected diffraction limit and an interesting reflection behavior for energy where the angle of incidence is not equal to the angle of reflection.  In many ways the magnetic system is much more exciting because the external magnetic field offers the opportunity to tune the dispersion relations and alter the focusing patterns, something that is not available in phonon focusing.

In this paper we study power flow from a point source in a ferromagnetic film with interfacial DMI.  In the ultrathin film limit and without DMI, the power flow is essentially isotropic, radiating energy approximately equally in all directions.  With DMI present however we find a set of remarkable results.  First, we show that a short pulse creates a bulls-eye pattern with a center that drifts away from the source over time.  Second, we find, both analytically and through micromagnetics, that with DMI one can create caustics, highly focused beams of energy, at particular frequencies. Finally, we find that a single point source, with DMI present, can create an interference pattern. The focusing patterns are highly nonreciprocal, with the caustic beams appearing only on one side of the film surface.  This has important implications for spintronic devices and applications, such as in magnonics, where the transfer of angular momentum and energy play a key role.

\begin{figure}
\includegraphics[width=8.5cm]{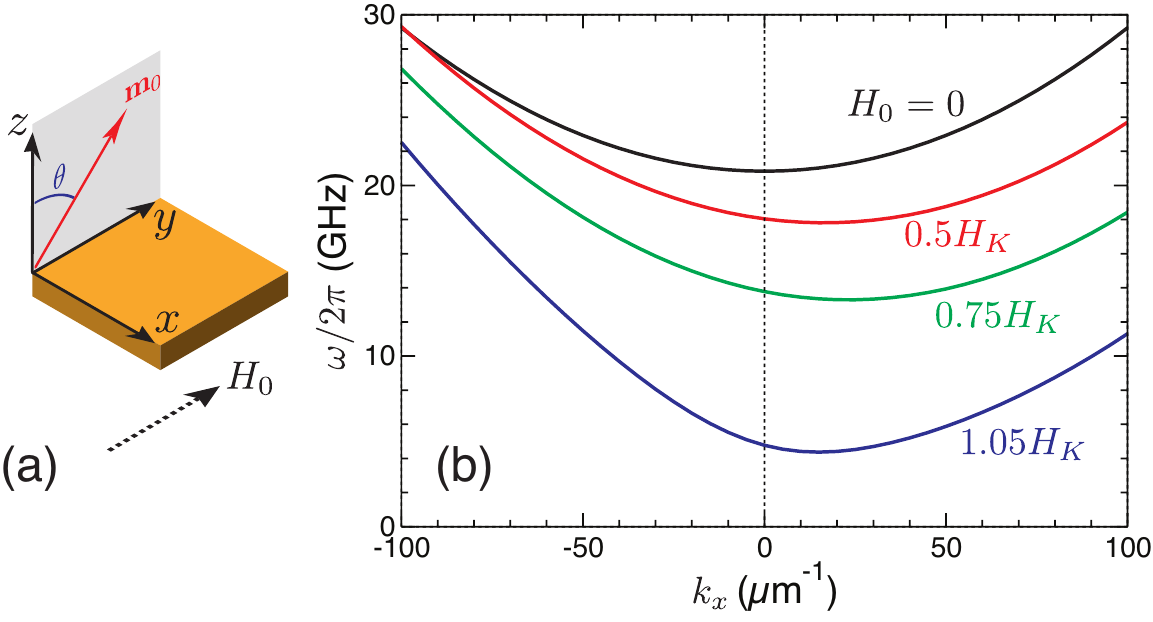}
\caption{(Color online) (a) Geometry of spin wave propagation. A magnetic field, $H_0$, is applied along $y$, which tilts the static magnetization by an angle $\theta$ away from the uniaxial anisotropy axis ($z$). (b) Dispersion relation ($k_y = 0$) for different $H_0$, with $D = 1$ mJ/m$^2$, based on Eqs. (\ref{eq:disptilt}) and (\ref{eq:dispsat}). $H_K$ denotes the anisotropy field.
}
\label{fig:geometry}
\end{figure}

Many of the features involving the nonreciprocity can be deduced from the spin wave dispersion relation~\cite{Moon:2013dm, Kostylev:2014gi}. We consider an interfacial DMI, which primarily involves ultrathin ferromagnets in asymmetric trilayers such as Pt/Co/Al$_2$O$_3$, Pt/Co/Ir, etc. Let $\mathbf{m} = \mathbf{m}_0 + \delta \mathbf{m}$ represent the magnetization and $\mathbf{H}_{\rm eff} = \mathbf{H}_{{\rm eff},0} + \delta \mathbf{H}_{\rm eff}$ the effective field, where $\mathbf{m}_0$ and $\mathbf{H}_{{\rm eff},0}$ are the static components and $\delta \mathbf{m}$ and $\delta \mathbf{H}_{\rm eff}$ are the dynamic components. The dispersion relation is obtained by linearizing the Landau-Lifshitz equation about the equilibrium state, $d \mathbf{m}/dt = -\gamma \mu_0 \left( \mathbf{m}_0 \times \delta \mathbf{H}_{\rm eff} + \delta \mathbf{m} \times \mathbf{H}_{{\rm eff},0} \right)$, where $\gamma$ is the gyromagnetic constant. The effective field comprises contributions from the exchange, perpendicular magnetic anisotropy along the $z$ axis, interfacial DMI, and the Zeeman energy associated with the applied magnetic field, $H_0 \hat{\mathbf{y}}$. The system geometry is illustrated in Fig.~\ref{fig:geometry}(a). 
For $H_0$ lower than the anisotropy field, $H_K = 2 K_0/\mu_{0} M_{s}$, where $K_0$ is the effective uniaxial anisotropy constant, $K_0 = K_u - \mu_0 N_z  M_s^2/2$, and $M_s$ is the saturation magnetization, $\mathbf{m}_0$ is tilted away from the film normal by an angle $\theta= \sin^{-1}\left(h \right)$, where $h \equiv H_0 / H_K \leq 1$. Here, $N_z = 1$ represents the demagnetization coefficient of an infinite thin film and $K_u$ is the strength of the interface-driven perpendicular magnetic anisotropy. The dispersion relation for this configuration is given by
\begin{equation}
\omega = \sqrt{ \left[ \omega_K + \omega_{\rm ex}(k) \right] \left[ \omega_K \left(1 - h^2 \right) + \omega_{\rm ex}(k) \right]   }  - \frac{2 \gamma D}{M_s} h k_x,
\label{eq:disptilt}
\end{equation}
where $\omega_K \equiv \gamma \mu_0 H_K$, $\omega_{\rm ex}(k) \equiv 2\gamma A k^2 / M_s$, and $k \equiv \| \mathbf{k} \|$. $A$ is the exchange and $D$ is the DMI constant. For $H_0 \geq H_K$, $\mathbf{m}_0$ is along $\hat{\mathbf{y}}$ and $\theta =\pi/2$. This leads to
\begin{equation}
\omega = \sqrt{ \left[ \omega_0 + \omega_{\rm ex}(k) \right] \left[ \omega_0 - \omega_K + \omega_{\rm ex}(k) \right]   }  - \frac{2 \gamma D}{M_s} k_x,
\label{eq:dispsat}
\end{equation}
where $\omega_0 \equiv \gamma \mu_0 H_0$.

Examples of $\omega(k)$ are shown in Fig.~\ref{fig:geometry} for several $H_0$. Under zero field, we observe a symmetric curve about $k_x = 0$, which indicates reciprocal propagation. Propagation is always reciprocal along $y$ in this geometry. As $H_0$ is increased and $\mathbf{m}_0$ tilts toward the film plane, the dispersion relation is displaced along the $k_x$ axis, which indicates nonreciprocal propagation. This displacement is largest when $H_0 \geq H_K$, as described by the linear $k_x$ terms in Eqs.~\ref{eq:disptilt} and \ref{eq:dispsat}. Indeed, it is this Damon-Eshbach geometry that has allowed the DMI strength to be probed in recent experiments~\cite{Di:2015fe, Di:2015fs, Nembach:2015ep, Belmeguenai:2015hj, Cho:2015eq}. In Fig.~\ref{fig:geometry}, we used parameters representative of ultrathin ferromagnetic films with perpendicular magnetic anisotropy, namely $A = 15$ pJ/m, $M_s = 1$ MA/m, $K_u = 1$ MJ/m$^3$, and $D = 1$ mJ/m$^2$.

\begin{figure}
\includegraphics[width=8.5cm]{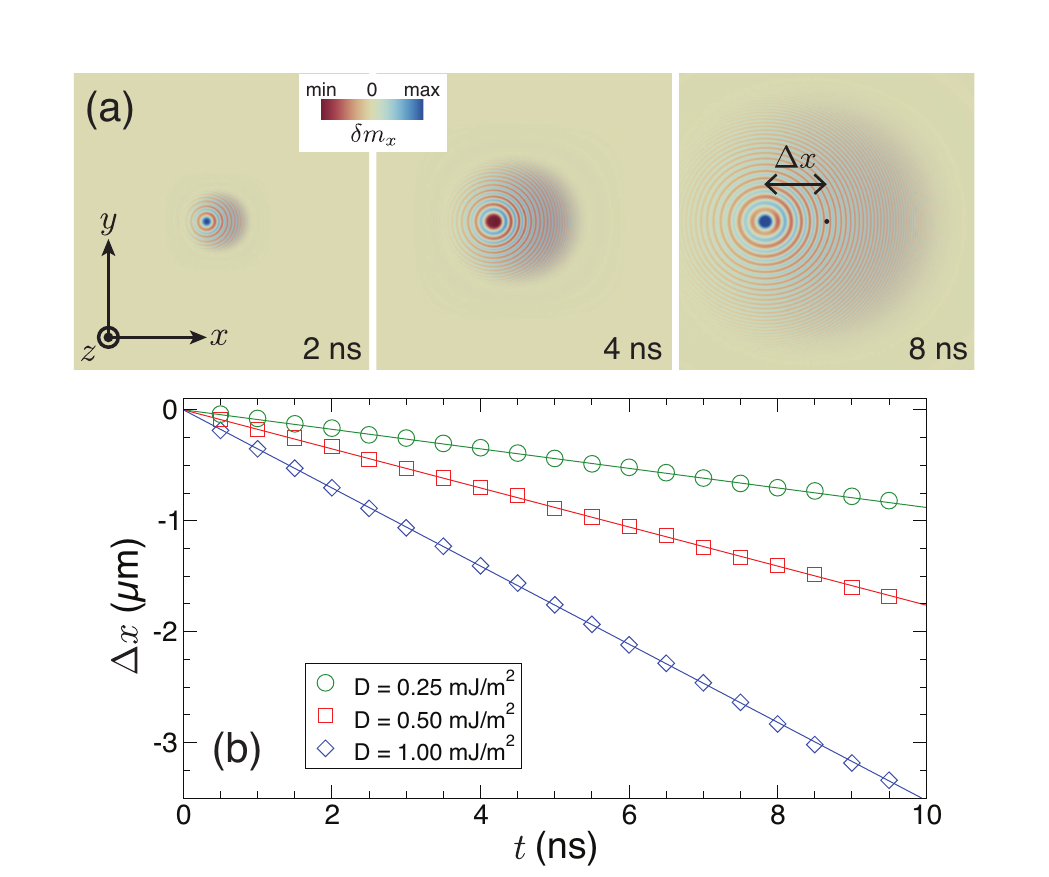}
\caption{(Color online) DMI-induced drift of a spin wave ripple. (a) Time evolution of the ripple 2, 4, and 6 ns after a sinusoidal field pulse at the image center ($D = 1$ mJ/m$^2$). The image dimensions are 10 $\mu$m $\times$ 10 $\mu$m. $\Delta x$ denotes the displacement of the ripple center. (b) Ripple displacement as a function of time for three $D$ values. Symbols represent simulation data while solid lines are based on Eq. (\ref{eq:dispsat}).}
\label{fig:bullseye}
\end{figure}

An interesting consequence of the shifted dispersion relation is shown in Fig.~\ref{fig:bullseye}, where we present results of micromagnetics simulations of the transient magnetic response to a pulsed field. 
We used the MuMax3 code~\cite{Vansteenkiste:2014et} and considered a 40 $\mu$m $\times$ 40 $\mu$m $\times$ 1 nm film that was discretized using $4096\times 4096 \times 1$ finite difference cells. (The smallest wavelength considered is $\sim 250$ nm, a value much larger than the cell size of $\sim 9.8$ nm.) We considered $\mu_0 H_0 = 0.8$ T ($\simeq 1.05 H_K$) and computed $\mathbf{m}(t)$ in response to a 5 GHz sinusoidal field excitation of 50 mT in amplitude along $\hat{\mathbf{x}}$ that was applied for one period (0.2 ns). The response comprises a ripple structure that represents spin waves radiating outward from the excitation source. For $D \neq 0$ the ripple center drifts along $-\hat{\mathbf{x}}$ as its size grows [Fig.~\ref{fig:bullseye}(a)]. In Fig.~\ref{fig:bullseye}(b), the ripple displacement is shown as a function of time for different $D$. The drift velocity of the ripple depends on $D$, where the lines indicate the expected displacement given by $v_{\rm drift} = \partial\omega_{\rm drift}/\partial k_x = \omega_{\rm drift}/k_x = -2 \gamma D/M_s$, which represents the component of $\omega(k)$ for which the phase and group velocities are identical. The DMI therefore conduces an underlying drift in the spin wave flow, which can be interpreted as a Doppler shift induced by an intrinsic spin current~\cite{Kikuchi:2016}.

\begin{figure}
\includegraphics[width=8.5cm]{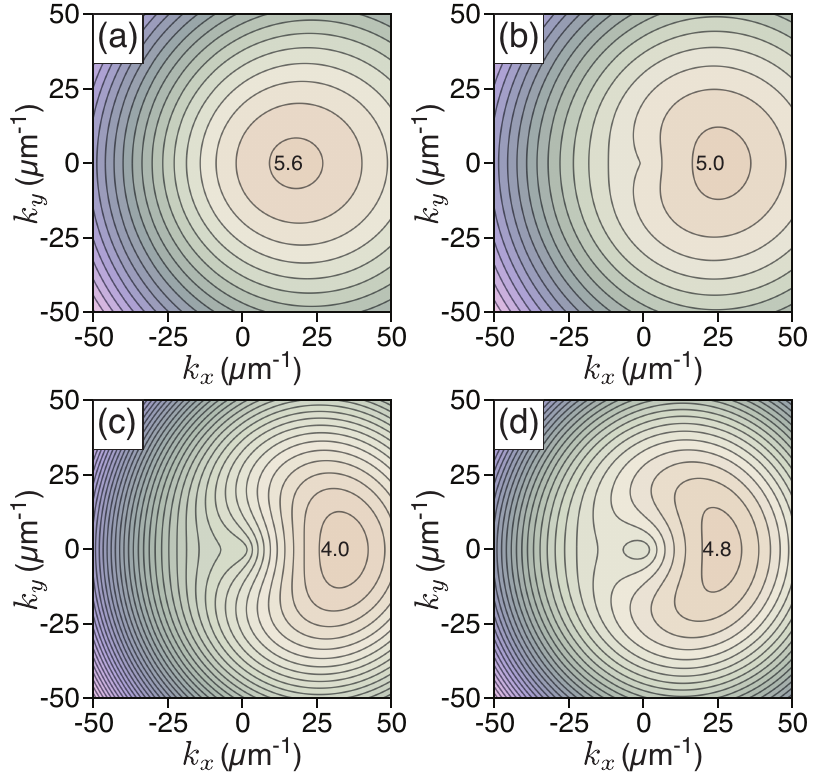}
\caption{(Color online) Frequency contours of Eq. (\ref{eq:dispdipole}) with $\mu_0 H_0 = 0.8$ T. $D = 1$ mJ/m$^2$ for (a) $d = 0$ nm, (b) $d = 1$ nm, and (c) $d = 2$ nm. (d) $D = 0.5$ mJ/m$^2$ and $d = 2$ nm. The lowest frequency contour is indicated (in GHz) and each successive contour represents a frequency difference of 0.2 GHz.}
\label{fig:freqcontour}
\end{figure}

We now discuss how this drift leads to focusing and caustics. The far-field radiation pattern of waves excited by a point source can be predicted from the slowness surface, i.e., a constant frequency curve in $k-$space. The radiation or focusing pattern can then be determined from the power flow, directed along the normal to the slowness surface, with an amplitude that is inversely proportional to the square root of the curvature of the slowness surface~\cite{Veerakumar:2006gy}. Caustics appear at points along the slowness surface at which its curvature goes to zero, resulting in a divergence in the power flow. To understand how caustics appear for spin waves in the ultrathin film, we return to the dispersion relation in Eq.~\ref{eq:dispsat}.  This is shown in Fig.~\ref{fig:freqcontour}(a), where each contour represents a slowness surface. 
While the contours are shifted from the origin in $k$-space for $D\neq 0$, the curvature is finite and positive everywhere since the contours remain largely circular by virtue of the exchange term, $\omega_{\rm ex} \propto A k^2$. We now consider the influence of the dipole-dipole interaction, which in the ultrathin film limit can be approximated by a local interaction in the following way~\cite{Arias:1999gy},
\begin{equation}
\omega(\mathbf{k}) = \sqrt{ \omega_{||}(\mathbf{k}) \, \omega_{\perp}(\mathbf{k})  }  - \frac{2 \gamma D}{M_s} k_x,
\label{eq:dispdipole}
\end{equation}
where $\omega_{||}(\mathbf{k}) = \omega_0 + \omega_{\rm ex}(k) + \gamma \mu_0 M_s d k_x^2/2\| \mathbf{k} \|$, $\omega_{\perp}(\mathbf{k}) = \omega_0 + \omega_{\rm ex}(k) - \omega_K  - \gamma \mu_0 M_s d \| \mathbf{k} \|/2$, and $d$ is the film thickness. In Figs.~\ref{fig:freqcontour}(b)-(d), we illustrate how the slowness surfaces change as the film thickness is increased and the dipolar interaction becomes more important. We can observe that a ``dent'' long the $-k_x$ axis appears for low frequencies, which is quite pronounced in Fig.~\ref{fig:freqcontour}(c). Moreover, a smaller value of the DMI ($D = 0.5$ mJ/m$^2$) for a 2-nm-thick film results in the appearance of a second slowness surface enclosed within the first [5.6 GHz contours, Fig.~\ref{fig:freqcontour}(d)]; we will revisit this point later. Importantly, the presence of the dent indicates that the curvature of the slowness surface changes sign, which means that caustics are created.

\begin{figure*}
\centering\includegraphics[width=17cm]{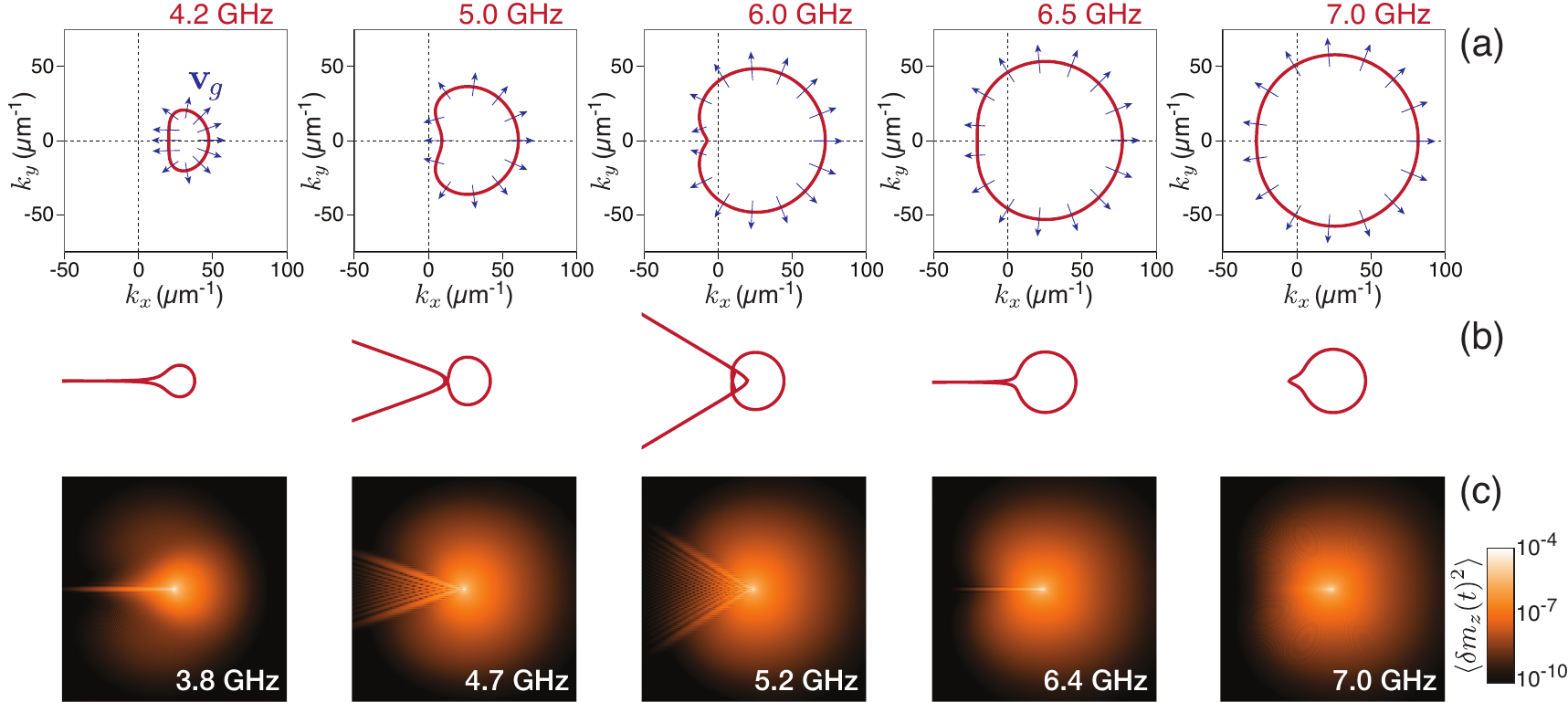}
\caption{(Color online) Spin wave focusing for $D = 1$ mJ/m$^2$ and $d = 2$ nm. (a) Slowness surfaces for different frequencies determined from Eq. (\ref{eq:dispdipole}). $\mathbf{v}_g$ denotes the group velocity vector. (b) Predicted focusing patterns based on (a). (c) Simulated focusing patterns due to a sinusoidal point source excitation at different frequencies. Each image represents an area of 20 $\mu$m $\times$ 20 $\mu$m with the point source at the center.
}
\label{fig:focusing}
\end{figure*}

Focusing patterns for $D = 1.0$ mJ/m$^2$ and $d = 2$ nm are shown in Fig.~\ref{fig:focusing}. 
We consider five different frequencies with distinct slowness surfaces [Fig.~\ref{fig:focusing}(a)]. The group velocity is indicated along each slowness surface. The expected focusing patterns are shown in Fig.~\ref{fig:focusing}(b), computed from the the curvature of the slowness surface in Fig.~\ref{fig:focusing}(a). For $\omega/2\pi = 4.2$ GHz, a caustic can be seen for propagation along $-x$, which results from the flattening on the left part of the slowness surface. As the frequency is increased to 5 and 6 GHz, a dent develops in the slowness surface, leading to two caustics propagating outward in the $-x$ direction. The dent leads to the curvature vanishing at two points along the slowness surface, resulting in the two focused beams predicted. As the frequency is further increased, the dent vanishes and a single caustic is recovered at 6.5 GHz. For higher frequencies, the exchange terms become dominant and the slowness surfaces recover a more elliptical shape, resulting in weaker focusing effects as seen for 7.0 GHz.

This behavior was reproduced in micromagnetics simulations, where the spin wave power flow from a point source excitation was computed. Using the geometry in Fig.~\ref{fig:bullseye}, we computed the response to a continuous sinusoidal point source field excitation at the center of the simulation grid. In Fig.~\ref{fig:focusing}(c), the spin wave power is presented for five excitation frequencies, which is computed by averaging the $z$ component of the dynamic magnetization, $\langle \delta m_z(\mathbf{r},t)^2 \rangle$, over two periods after 150 periods of the field excitation. The excitation frequencies used in the simulations were chosen to match as closely as possible the focusing patterns predicted from the dispersion relation [Fig.~\ref{fig:focusing}(b)]. While the agreement in the frequencies is only semi-quantitative, the simulations reproduce well the different focusing patterns predicted, namely the orientation and trends in the different caustics as the excitation frequency is increased. The discrepancy is likely due to the local approximation used for the dipolar interaction in Eq.~\ref{eq:dispdipole}. Nevertheless, there is a good agreement between the theory and simulation.

\begin{figure}
\includegraphics[width=8cm]{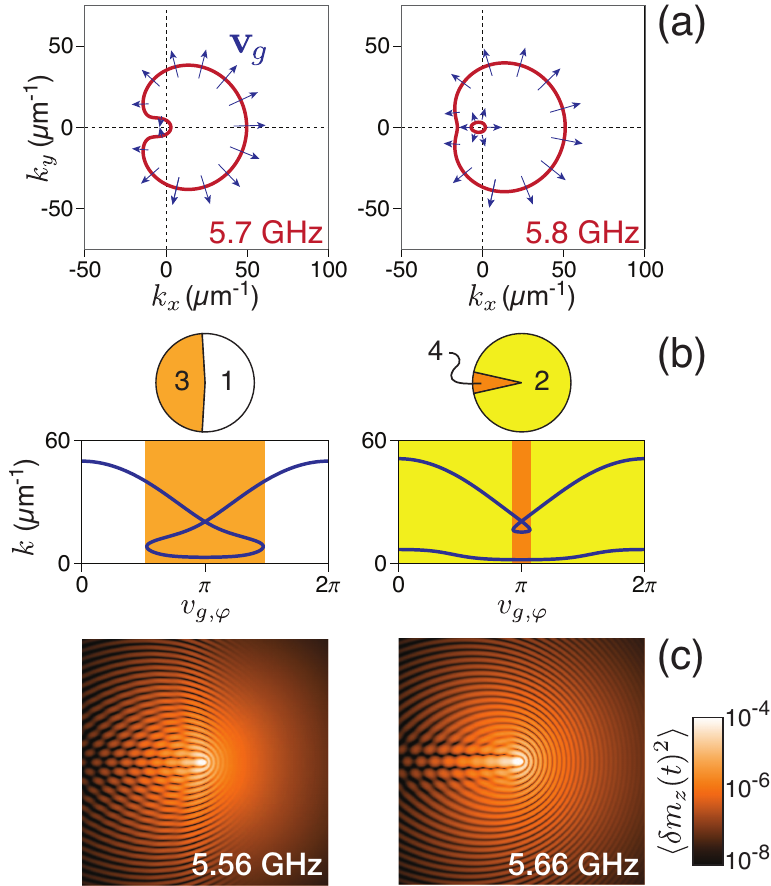}
\caption{(Color online) Interference patterns for $D = 0.5$ mJ/m$^2$ and $d = 2$ nm. (a) Slowness surfaces based on from Eq. (\ref{eq:dispdipole}). $\mathbf{v}_g$ denotes the group velocity vector. (b) $k = \| \mathbf{k} \| $ as a function of the $\mathbf{v}_g$ orientation for the slowness surfaces in (a). The shaded regions denote propagation directions for which several $k$ are possible. In the top inset, propagation directions along which interference is expected are indicated, where the number of allowed $k$ are shown. (c) Simulated interference patterns due to a point source excitation at different frequencies. Each image represents an area of 5 $\mu$m $\times$ 5 $\mu$m with the point source located at the center. The frequencies are chosen to match the interference patterns expected from (b).}
\label{fig:interference}
\end{figure}

Another remarkable feature of Eq. (\ref{eq:dispdipole}) is the possibility of generating interference patterns from a single point source. Some evidence of interference can already by seen in Fig. \ref{fig:focusing}(c) for 4.7 and 5.2 GHz in the region bounded by the two focused beams. To see how interference arises, consider the case of $D = 0.5$ mJ/m$^2$ and $d = 2$ nm [Fig. \ref{fig:freqcontour}(d)] for which the dent in the slowness surface evolves into two distinct surfaces between 5.7 and 5.8 GHz, as shown in Fig.~\ref{fig:interference}(a). 
Consider the response at 5.7 GHz, which results in a C-shaped slowness surface. If we examine how the group velocity vector, $\mathbf{v}_g$, evolves around this surface, we notice that certain orientations of $\mathbf{v}_g$ appear at multiple points along this surface, which indicates that propagation along these directions involve partial waves with different $\mathbf{k}$. To see this, we plot in Fig.~\ref{fig:interference}(b) $k = \| \mathbf{k}\|$ as a function of the angle of $\mathbf{v}_g$ with respect to the $k_x$ axis (in the film plane), $v_{g,\phi}$, for the two excitation frequencies considered. For 5.7 GHz, three $k$ are allowed over a range of propagation angles, while only a single $k$ is allowed elsewhere [top inset of Fig.~\ref{fig:interference}(b)], which suggests three-wave interference should occur for propagation near the $-x$ direction, while no interference is expected along $+x$. This was verified with micromagnetics at a similar frequency of 5.56 GHz, where interference is mostly localized to the $x < 0$ region. On this basis,  the existence of two slowness surfaces for 5.8 GHz [Fig.~\ref{fig:interference}(a)] should result in interference for all propagation directions; we find that four-wave interference is expected within a narrow range of propagation angles about the $-x$ direction, while two-wave interference for all other directions [Fig.~\ref{fig:interference}(b)]. This was also confirmed in simulation at 5.66 GHz, where two different interference patterns with the expected angular dependence can be seen.

Our results suggest that similar effects can appear in thicker films with spin-polarized currents. Since the DMI induces an overall drift in the spin wave flow (Fig.~\ref{fig:bullseye}), analogous effects should arise with other mechanisms that induce a drift, such as spin transfer torques~\cite{Vlaminck:2008ij}. In this case, a spin current drift velocity of $u = J P \hbar \gamma / (2 e M_s)$ is generated, where $J$ is the current density and $P$ is the spin polarization. We have verified this using micromagnetics, where identical results to Fig.~\ref{fig:focusing}(c) were obtained with $D = 0$ but instead with a uniform current density of $J = 6.08$ TA/m$^2$ ($P = 1$) along $\hat{\mathbf{x}}$, which results in the same drift velocity as the DMI-induced value of $v_{\rm drift} = $ 352.2 m/s with $D = 1$ mJ/m$^2$. Note that such focusing effects are not confined to thin films with perpendicular magnetic anisotropy but should also appear in planar systems provided an underlying spin-wave drift is present.

Magnetostatic nonreciprocity, used in microwave circulators and isolators~\cite{Stancil:2009gc}, generally requires 1-50 $\mu$m-thick films.  In contrast the nonreciprocity seen here is found in nm-thick films.  The ability to control caustics and interference patterns in thin films might also find use in microwave devices such as demultiplexers~\cite{Gruszecki:2016hr}, band pass filters, and isolators. The caustic beams could also be useful in magnon-based logic circuits~\cite{Khitun:2010dx}, holographic devices~\cite{Khitun:2013hz}, and for exploring magnetic analogs of wave phenomena seen in other physical systems such as electron optics~\cite{Petersen:2013ik} and phonons~\cite{Every:2003fe}.

%%
%	end text
%%

%%%
%	Acknowledgements
%%%
\begin{acknowledgments}
The authors acknowledge fruitful discussions with F. Garcia-Sanchez. This work was partially supported by the Agence Nationale de la Recherche (France) under Contract No. ANR-14-CE26-0012 (Ultrasky).
\end{acknowledgments}

%%
% 		References
%%
\bibliography{articles}

\end{document}